Enhancing Retrieval Processes for Language Generation with Augmented Queries

Julien Pierre Edmond Ghali, Kosuke Shima, Koichi Moriyama, Atsuko Mutoh, Nobuhiro Inuzuka

Nagoya Institute of Technology, Japan

Corresponding author: Julien Pierre Edmond Ghali, julienpe.ghali@gmail.com

2## Abstract

In the rapidly changing world of smart technology, searching for documents has become more challenging due to the rise of advanced language models. These models sometimes face difficulties, like providing inaccurate information, commonly known as "hallucination." This research focuses on addressing this issue through Retrieval-Augmented Generation (RAG), a technique that guides models to give accurate responses based on real facts. To overcome scalability issues, the study explores connecting user queries with sophisticated language models such as BERT and Orca2, using an innovative query optimization process. The study unfolds in three scenarios: first, without RAG, second, without additional assistance, and finally, with extra help. Choosing the compact yet efficient Orca2 7B model demonstrates a smart use of computing resources. The empirical results indicate a significant improvement in the initial language model's performance under RAG, particularly when assisted with prompts augmenters. Consistency in document retrieval across different encodings highlights the effectiveness of using language model-generated queries. The introduction of UMAP for BERT further simplifies document retrieval while maintaining strong results.

Material and methods: the code is available at this link: https://github.com/JulienGha/RAGDoc2Vec. Due to copyrights reason, the data used are not available, but users can try with their own data.

# 1. Introduction

In the field of language processing and machine learning, the quest for precision and contextual understanding in document retrieval has become more crucial than ever to address the phenomenon of "hallucination" [1], referring to the generation of inaccurate, non-sensical, or detached text, that occurs when using large language models (for example when an output includes a reference that doesn't exists). At the heart of this quest is the concept of Retrieval-Augmented Generation (RAG) [2], a method of providing a model with factual knowledge and basing its response on it. However, this system may encounter difficulties in terms of scalability, or its ability to find the right documents; the size of the input message given to our model being limited, it is not possible to simply provide it with a complete book on the subject and simply prompt it to use only the relevant parts. In this article, we present a query augmentation model that could enhance the way RAG models generate responses.

Query Augmentation represents a strategic alignment of user queries during document retrieval with the advanced capabilities of language models such as T5 [3], BERT [4], GPT [5] or Orca2 [6]. In our research, the goal was to generate, from an initial user query, a response likely to be found in a document. Our intuition was that in an environment using document and query vectorization, it was more likely to retrieve the desired information with similar data rather than through a question or question/answer training. In the context of augmented generation for our research, query augmentation is not limited to providing the right answers; it's about designing queries that guide the AI model to understand and reflect the semantic nuances of the user's demand, ultimately leading to more relevant and contextually appropriate document retrieval.

By enhancing the initial steps of the retrieval process through augmented queries, we enable AI models to sift through extensive data repositories with greater accuracy and relevance. This approach not only elevates the efficiency of document retrieval systems but also enriches the user experience by ensuring that the information retrieved closely matches their search intent.

## 2. Related works

Natural Language Processing (NLP) has witnessed significant advancements with the emergence of LLMs. Despite their prowess in downstream tasks, these models face challenges in accessing and manipulating knowledge, leading to suboptimal performance on knowledge-intensive tasks. A groundbreaking approach to address these limitations is Retrieval-Augmented Generation (RAG). In Lewis et al.'s seminal work (2020) [2], the authors introduce RAG models that combine pre-trained parametric and non-parametric memory for language generation. By leveraging a pre-trained seq2seq model as parametric memory and a dense vector index of Wikipedia as non-parametric memory, RAG achieves superior performance on a spectrum of knowledge-intensive NLP tasks. It outshines parametric seq2seq models and task-specific architectures, setting new benchmarks in open-domain question answering (QA) tasks. Furthermore, RAG models exhibit enhanced language generation capabilities, producing more specific, diverse, and factual language compared to state-of-the-art parametric-only seq2seq baselines.

In the realm of information retrieval (IR), recent research has made remarkable strides. Hambarde and Proenca (2023) [7] provide an extensive overview of IR models, discussing the state-of-the-art methods, including those based on terms, semantic retrieval, and neural approaches. Karpukhin et al. (2020) [8] revolutionize open-domain QA by introducing dense passage retrieval, demonstrating its practical implementation using only dense representations.

This approach outperforms traditional sparse vector space models, leading to state-of-the-art performance on multiple open-domain QA benchmarks. Qu et al. (2020) [9] further optimize dense passage retrieval with RocketQA, addressing challenges such as training-inference discrepancy and limited training data. RocketQA significantly outperforms previous state-of-the-art models on well-established datasets, showcasing its effectiveness in improving end-to-end QA systems. Advancements in similarity search, as demonstrated by Johnson et al.'s (2019) [10], their work, optimizing k-selection and addressing memory hierarchy challenges, sets new benchmarks in various similarity search scenarios. The ability to construct a high-accuracy k-NN graph on a massive dataset underscores the ongoing effort to enhance scalability and efficiency in information retrieval systems, this system is better known as FAISS.

The evolution of text representation methods has been pivotal in advancing NLP capabilities. Le and Mikolov's Doc2Vec algorithm (2014) [11] overcomes the limitations of bag-of-words models by learning distributed representations of sentences and documents. This approach outperforms traditional models in text classification and sentiment analysis tasks, achieving new state-of-the-art results. Devlin et al.'s BERT (2018) introduces bidirectional pre-training, empowering the model to achieve state-of-the-art results across various NLP tasks. These methods lay the foundation for effective text representation, enabling models to capture complex semantic relationships within language.

Advancements in query augmentation strategies play a crucial role in improving information retrieval effectiveness. Qiu and Frei (1993) [12] propose a probabilistic query expansion model based on a similarity thesaurus, demonstrating notable improvements in retrieval effectiveness. Voorhees (1994) [13] explores lexical query expansion using WordNet synonym sets, showing the utility of expanding queries based on lexical-semantic relations. Bai et al. (2007) [14] advocate for query-specific contexts, integrating both contexts around and within the query to enhance retrieval effectiveness. These strategies highlight the significance of contextual information in refining the search process and delivering more relevant results.

Lastly, prompt optimization techniques contribute to the refinement of large language models. Chung et al. (2022) [3] scale instruction-finetuned language models, showcasing substantial performance gains across various benchmarks. Zheng et al. (2023) [15] introduce Step-Back Prompting, a technique that enables large language models to perform abstractions by deriving high-level concepts from specific instances. This approach significantly improves the models' reasoning abilities, marking a substantial leap in their performance on reasoning-intensive tasks.

In conclusion, the intersection of retrieval-augmented generation, information retrieval, text representation, query augmentation, and prompt optimization is a cornerstone in the current landscape of NLP research. Leveraging these advancements provides a robust foundation for augmenting query processes, offering the potential to enhance the precision, diversity, and factual accuracy of natural language understanding systems.

### 3. Our Contribution

This article introduces a novel approach to query augmentation in the context of document retrieval, aiming to enhance the efficacy of retrieval-augmented generation systems. The traditional paradigm of relying solely on user queries for document retrieval, or on pairs of query/answer is challenged by the integration of a prompt augmenter, designed to dynamically analyze lexical fields, and generate augmented search queries. By harnessing the power of Language Models (LMs), our methodology provides a nuanced understanding of user intent and significantly enhancing the precision and relevance of retrieved documents.

# 4. Methods

## 4.1 Methodology

Regarding the optimization of document retrieval, our methodology rests on the hypothesis that the insertion of an intermediate document, strategically generated between the user's query and the retriever, could yield enhanced efficiency compared to relying solely on the user's query or a learned pattern of question/answer. Traditional document retrieval systems often hinge on the precision of user queries, expecting these queries to perfectly encapsulate the user's information needs. However, such an approach may encounter challenges when user intent is nuanced or when the available documents possess varying degrees of contextual relevance.

To address this, we propose the integration of a generator as a methodological step between the user's query and the retriever, as shown in the **Fig 1**. This intermediate document, generated dynamically and semantically aligned with the user's query, acts as a contextual bridge. Its purpose is to provide the subsequent retriever with a more nuanced representation of the user's intent, thereby refining the search process. Our methodological approach is founded on the belief that this generated document, shaped by advanced language models, can capture subtle nuances and context that may be overlooked in the original user query. The aim is to augment the precision and efficiency of the document retrieval process, ensuring that the subsequent retriever is armed with a more contextually rich input for its search.

The potential benefits of this method are diverse. It not only offers a mechanism to handle vague or ambiguous queries but also provides adaptability to users with varied communication styles. Additionally, the generated document serves as a catalyst for refining the retriever's search, offering guidance for a more targeted exploration of the document corpus. In the following sections, we delve into the practical implementation of this method, exploring the role of prompt optimization and document generation in tandem with retrieval-augmented models. This method aims to bridge the gap between user queries and retriever algorithms, testing the efficacy of our hypothesis in optimizing the document retrieval process for greater efficiency and precision.

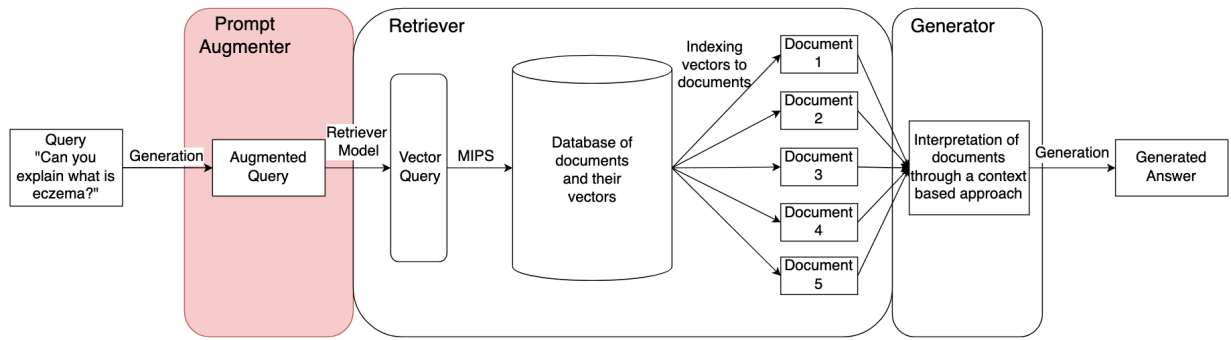

Fig 1. Proposed system architecture

**4.2 Prompt Augmenter**

The Prompt Augmenter operates as a query augmentation tool utilizing a language model generator. The process involves guiding users to distill their original prompts into the underlying questions they seek to answer. Subsequently, the generator, trained to comprehend and generate human-like text, transforms the user's prompt into an augmented query. This augmented query is carefully crafted to mirror the language commonly found in documents related to the user's inquiry.

By employing a language model generator for query augmentation, the system facilitates a more refined and contextually aligned interaction with document retrieval. The generated augmented queries serve as effective probes, enhancing the precision of the retrieval process. The interest in utilizing this approach lies in its ability to bridge the gap between user input and the language structures prevalent in relevant documents, ultimately leading to more accurate and relevant document retrieval results.

**4.3 Retrievers**

We decided to use word embedding methods due to the nature of our Prompt Augmenter; we were looking for similar documents with a document.

- TF-IDF: time complexity: o (n * m)
    - Where 'n' represents the number of documents, 'm' is the average number of unique terms in a document.
- Doc2Vec: time complexity: o (n * d * e + n * n * d)
    - Where 'n' is the number of documents, 'd' is the dimensionality of the document vectors, and 'e' is the number of training epochs for Doc2Vec, we used 400. The first term represents the training complexity, and the second term represents the retrieval complexity for calculating cosine similarity between the query vector and all document vectors.
- BERT: time complexity o (n * e + n * $d^2$) when using dimension reduction
    - Where n is the number of documents, e is the time complexity of encoding a single document with BERT, d is the number of dimensions in the UMAP space, it scales quadratically with the amount of dimension, and we decided to only use 2 here. The time complexity for BERT encoding is o (n·e), and for UMAP, it's approximately o (n·d).
    - If we weren't using dimension reduction, then the complexity would be the following: o (n·l·h·h′+n·n·h′).

- Where l is the maximum sequence length, h is the number of attention heads, and H' is the hidden size. This complexity considers the encoding of each document o(n·l·h·h′) and the pairwise similarity calculation o(n·n·h′).

We highlighted the different time complexities for the retrieval each method, but there is to consider the precision to evaluate its efficiency, as well as the training time. While some documents are fast to encode, like TF-IDF, some takes longer time, which is the case for BERT, that ended being 44 times slower.

In the context of our research, we implemented a strategic approach to enhance document retrieval. When using PDF, we broke it down into sentences, focusing on those longer than 15 characters. During the retrieval process, we deliberately selected not only the targeted sentence (n) but also the preceding (n-1) and succeeding (n+1) sentences. This decision assumed that relevant information and contextual clues are often dispersed before and after the identified sentence, contributing to a more comprehensive understanding of the content.

**4.4 Generator**

In our framework, the generator functions as an answer generation tool centered around documents, powered by a language model generator. When presented with a user query and a context formed by retrieved documents, the system directs the model through a systematic procedure. This involves instructing the model to craft a response to the user's query within specific constraints, here the size of the answer due to computation constraints and the, prompting the incorporation of pertinent knowledge from the provided context.

Throughout this process, the model generates a response by tapping into its trained knowledge base and considering the context extracted from documents. The anticipated outcome is a concise, relevant, and contextually embedded response that aligns with both the user's query and the information gleaned from the retrieved documents. This approach ensures that the model provides informative and contextually fitting answers, meeting the user's inquiry in coherence with the available document context.

## 5. Results

We applied our system to address a diverse set of 10 queries, following a systematic evaluation approach. Initially, we generated responses without incorporating Retrieval-Augmented Generation (RAG). After, we produced responses employing our three distinct retrieval methods but without our prompt augmentation technique, so just with the initial query. Finally, we utilized our three retrieval methods alongside prompt augmentation for response generation.

For the augmentation process, we selected Orca2 7b, a Small Language Model recognized for its outstanding performance compared to other Large Language Models. This choice was influenced by its impressive results combined with minimal GPU memory requirements, ensuring efficiency in our experimentation. As the source for retrieval, we first thought of using the Jeopardy dataset, we could have randomly selected the questions to measure our retriever capacity as well as our generator ability to understand the given context and generates answers, but we realized that the Jeopardy dataset only provides short answer, not a sentence, therefore if we wanted to efficiently measure our system we should have change the answer structure into full sentence (example: for the question "1912 Olympian; football star at Carlisle Indian School; 6 MLB seasons with the Reds, Giants & Braves", the answer should be "Jim Thorpe is a 1912 Olympian, a football star at Carlisle Indian School, he played 6 MLB seasons with the Reds, Giants & Braves" and not just only "Jim Thorpe"). Therefore, we decided to only use a book, "*The Cognitive Neuropsychology of Schizophrenia (Classic Edition)*" by

Frith, C. D. edition 2015. Later to test the efficiency of the information gain made by our system, we implemented another article dealing on the prevalence of alcohol use for persons having schizophrenia, wrote by Koskinen et al. [16] and an article explaining through experiment on diverse patients that paranoia, a symptom of schizophrenia, is the product of grandiosity and guilt, and can also be associated with bipolarity, written by Lake C. [17]. Employing resources like those underscores our system's capability to deliver fine-tuned results, by using RAG alone.

Regarding our evaluation metric and the intricacy of assessing Large Language Models (LLM) answers – if not ROUGE or BLEU for retrieval – we decided to propose our own based on a tree structure. We employed the logic outlined in Fig 2., signifying that a score for an answer is between 1 and 4, 4 being the highest score. If the answer is totally unrelated, as we can see sometimes when an output is unrelated training data, then the score is 1. If an answer is wrong, no matter the justification and how the system followed the procedure the score will only be 2. A correct answer but without properly respecting the constraints or the context is 3, while an answer that use the given context and respect the constrains (less than a specific number of words for example) will be given a 4.

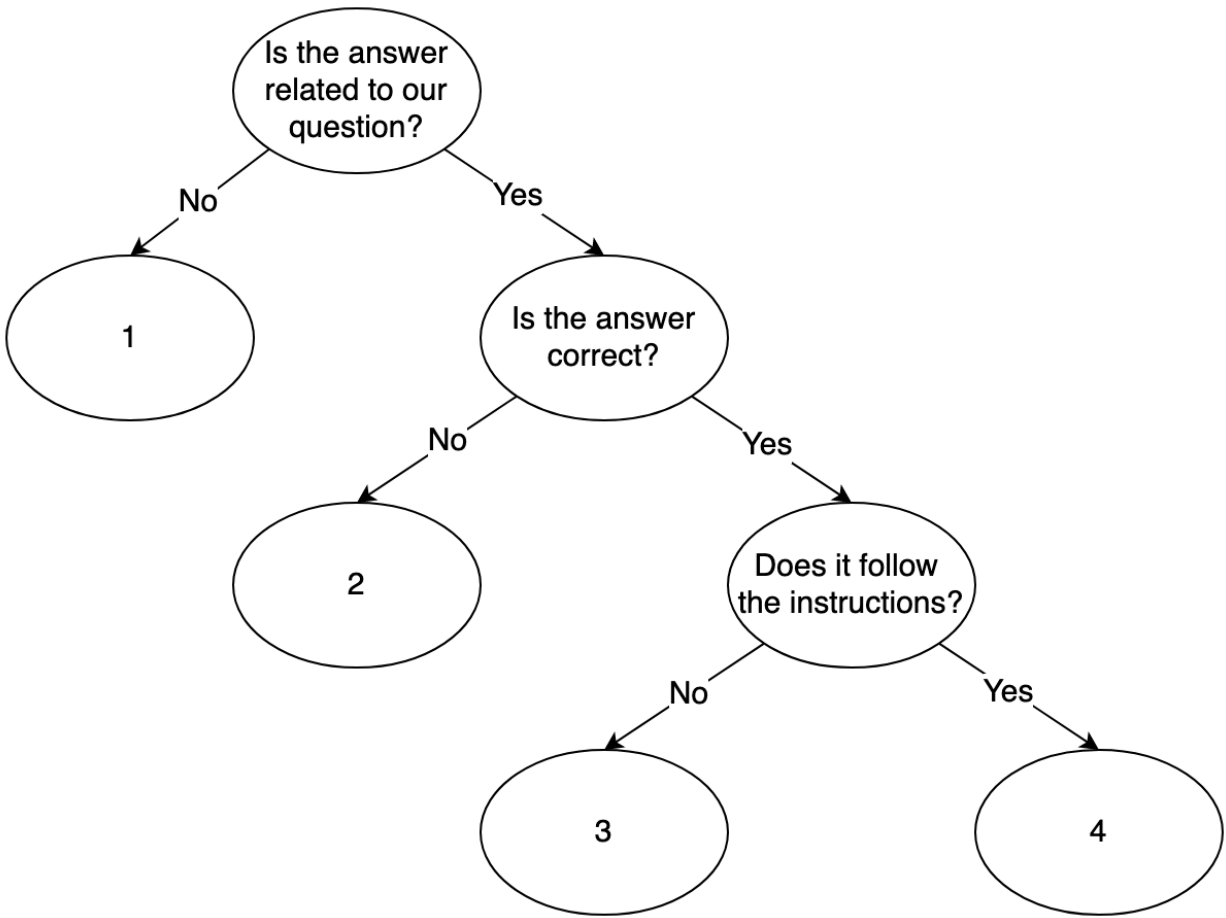

Figure 2. Answer evaluation system

To evaluate our different prompts, the results displayed on the **Tab 1.**, where for each system we put the of each evaluation made with our system.

|  | TF-IDF | Doc2Vec | BERT&UMAP |
|---|---|---|---|
| No RAG & No query augmentation | | 3 | |
| RAG & No query augmentation | 3.2 | 3.3 | 3.5 |
| RAG & query augmentation | 3.6 | 3.5 | 3.8 |

Table 1. Score based on retrieval methods

**Conclusion**

After carefully checking Tab 1. and the extra tables in Annex A, a clear improvement in the initial language model's performance is noticeable. RAG contributes to refining its capabilities, similar to fine-tuning. The prompt augmenter plays a crucial role in delivering factual responses, especially evident in the query "Can animals have schizophrenia," where the prompt augmenter stands out as the only method providing the correct answer.

The retriever capabilities showed consistency despite using different encodings, confirming our intuition about retrieving similar documents through those generated by our Language Model. The use of UMAP for BERT proves to be a successful strategy, simplifying retrieval complexity while maintaining consistent and reliable results, marking another achievement in our research.

By incorporating other resources, we aimed to fine-tune the results obtained solely through RAG. The efficiency of our system was further highlighted by introducing another article on the prevalence of alcohol use disorders in schizophrenia by Koskinen et al. The utilization of these additional articles showcased the capacity of our system to yield refined and information-enriched outcomes, affirming the significance of augmenting retrieval with well-curated and relevant documents. Those incorporations showed limited success; it also led to misunderstanding when the query lacks precision, it also exponentially increases the BERT retrieving time, which was expected when looking at the time complexity equations.

Overall, the performance difference with each system highlights their known strengths and weaknesses. BERT shows the best results, but the computation time is the highest, making it difficult to apply on a large-scale system. TF-IDF stands here as the best ratio performance/cost, while being the fastest it is also the one that benefit the most from our augmented query, being more reliable than Doc2Vec, due to its word counting nature. Doc2Vec, despite being performant, shows the weakest results from our query augmenter.

**Ethical Consideration**

In delving into the ethical considerations of our research, it's crucial to recognize the ongoing concerns tied to AI and language models. As these models get fancier, there's a growing worry about unintended consequences and ethical dilemmas. A big concern here is making sure these models are used responsibly, creating content that aligns with ethical norms and societal values.

Now, when we layer on Retrieval-Augmented Generation (RAG) and our innovative methods, it adds another ethical dimension. If the documents used to train the model raise a red flag – then there's a risk of people sneaking in harmful content. This could mean the model ends up spreading information that might hurt individuals or communities. The real challenge is that AI systems struggle to grasp the subtle differences between what's harmless and what might cause harm, a struggle we've seen in early versions of models like ChatGPT.

So, the ethical focus in our research is on building solid safeguards and oversight to tackle the risks tied to content generation and retrieval. Striking a balance between pushing the boundaries of innovation and using these models responsibly is key to ensuring our work has a positive impact on society.

## Limitations

Our research encountered constraints primarily stemming from hardware limitations, specifically the utilization of 8GB of GPU RAM. The processing time for a single query averaged around 10 minutes, particularly for generating answers spanning approximately 60 words. Additionally, constrained by both RAM and budget considerations, we opted for the utilization of local language models with limited capacity, exemplified by the Orca2 7B model. These limitations, while impacting the scale and speed of our query optimization process, served as crucial parameters that shaped the scope of our experimentation and analysis.

During the preparation of this work the author(s) used Chat GPT 3.5 in order to spell check and assist during writing. After using this tool/service, the authors reviewed and edited the content as needed and take full responsibility for the content of the publication.

This research did not receive any specific grant from funding agencies in the public, commercial, or not-for-profit sectors.

# Annexes

Annex A:

| Define hallucination | TF-IDF | Doc2Vec | BERT |
|---|---|---|---|
| No RAG & No query augmentation | Hallucinations are perceptions of sensory stimuli that occur in the absence of an external stimulus. They can involve any of the five senses: sight, hearing, taste, touch, or smell. Hallucinations can be caused by various factors, such as mental disorders, drug or alcohol use, sleep deprivation, or extreme stress. They can be vivid, realistic, and sometimes frightening, but are not considered delusions as they are not based on false beliefs. | | |
| RAG & No query augmentation | Hallucinations are sensory perceptions that occur without an external stimulus, often involving sight, sound, or touch. In the context of schizophrenia, auditory hallucinations are more likely to occur in situations where there is a discrepancy between the literal and intended meaning of speech, such as metaphors and irony. These hallucinations may be based on inner speech that the patient misattributes to an external source. The experience underlying many "auditory" hallucinations may occur at a more abstract level, without a sensory component. | Hallucinations are sensory perceptions that occur without external stimuli, often involving sight, sound, or touch. In the context of schizophrenia, auditory hallucinations are a common symptom, where individuals hear voices or sounds that are not present in reality. These hallucinations can be influenced by factors such as the complexity of auditory stimuli and the presence of random noise. In some cases, hallucinations may be related to inner speech, which is the continuous stream of thoughts and mental images experienced by an individual. Understanding the nature and causes of hallucinations is crucial for diagnosing and treating schizophrenia and other mental health conditions. | Hallucinations are perceptions that occur without any external stimulus, often associated with mental health conditions like schizophrenia. In the context provided, auditory hallucinations may be based on inner speech, which the patient misattributes to an external source. These experiences can be influenced by imbalances between high and low level processes in the brain, particularly in schizophrenic patients. |
| RAG & query augmentation | Hallucinations are perceptions that occur when an individual experiences sensory stimuli that are not present in reality. They are often attributed to intrusions of material from long-term memory, which the person mistakenly believes to be coming from an external source. In the context provided, hallucinations are observed in schizophrenia, where there is a lack of structure in memory and cognitive processing. This can lead to difficulties in distinguishing between self-generated and experimenter-generated items, as well as impairments in source memory and self-monitoring. | Hallucinations are sensory experiences that occur without any external stimulus, often associated with mental disorders like schizophrenia. In schizophrenia, hallucinations can manifest in various forms, such as auditory (hearing voices) or visual (seeing things). These hallucinations can be categorized as third-person (voices discussing the patient) or second-person (voices talking directly to the patient). Other symptoms of schizophrenia include thought insertion, thought broadcast, thought withdrawal, delusions of control, delusions of reference, and paranoid delusions, which are referred to as "negative" symptoms because they involve a reduction or absence of normal behaviors. These hallucinations and symptoms may be the result of a disorder in the brain, as suggested by the evidence presented in the context. | Hallucinations are sensory perceptions that occur without external stimuli, often involving sight, sound, or touch. In the context of schizophrenia, hallucinations can manifest as auditory, such as hearing one's own thoughts spoken aloud, or visual, like seeing people or objects that are not present. These experiences can be explained by cognitive defects, like a self-monitoring theory, which suggests that individuals with schizophrenia may struggle to distinguish between their own thoughts and reality. This theory can also explain certain types of auditory hallucinations, particularly when a person hears their own thoughts spoken aloud. However, it may not fully explain all forms of hallucinations, as some patients may falsely believe that many people are trying to communicate with them. |

| Can animals have schizophrenia | TF-IDF | Doc2Vec | BERT |
|---|---|---|---|
| No RAG & No query augmentation | Yes, animals can exhibit behavioral disorders similar to schizophrenia. In animals, these disorders are often referred to as "psychiatric" or "behavioral" disorders. Symptoms may include abnormal social behavior, repetitive or compulsive actions, and changes in normal patterns of behavior. However, the diagnosis and treatment of such disorders in animals may differ from those in humans due to differences in species-specific behaviors and communication. | | |
| RAG & No query augmentation | Animals, including humans, can experience mental disorders like schizophrenia. Schizophrenia is a complex disorder that affects an individual's thoughts, feelings, and behavior. It is characterized by symptoms such as hallucinations, delusions, and disorganized thinking. While animals do not have the same cognitive abilities as humans, they can exhibit some behaviors that may be related to schizophrenia. For example, vervet monkeys with neocortical lesions show changes in their social behavior, which could be similar to the effects of brain damage in schizophrenia. However, it is important to note that the exact causes and mechanisms of schizophrenia in humans and animals are still not fully understood, and more research is needed to determine if there is a direct connection between the two. | Yes, animals can have schizophrenia-like symptoms, as some signs of schizophrenia can be studied directly in animals. Stereotyped behavior has been extensively researched in animals, and there is evidence that the dopamine system is involved. Rats and monkeys have been used in experiments to study learning, memory, self-monitoring, and cognitive processes related to schizophrenia. However, there is still debate and disagreement among researchers about the significance of these findings and the exact nature of animal models for schizophrenia. | Animals do not have schizophrenia, as it is a human psychiatric disorder. The context provided discusses various aspects of schizophrenia, such as its etiology, movement disorders, and effects on cognition. However, these findings and research are specific to humans and do not apply to animals. |
| RAG & query augmentation | Animals, including non-human primates like chimpanzees and vervet monkeys, can exhibit behaviors related to social cognition and mental state perception. However, there is no direct evidence of schizophrenia in animals. Schizophrenia is a complex human psychiatric disorder characterized by disruptions in thought processes, perceptions, emotional responsiveness, and social interactions. While some animal behaviors may resemble aspects of schizophrenia, such as disrupted social cognition, these are not the same as the human condition. Further research is needed to understand the similarities and differences between animal behaviors and schizophrenia in humans. | Animals do not have schizophrenia, as it is a human psychiatric disorder. The context provided discusses hallucinations, brain abnormalities, and theories of schizophrenia, but it does not mention animals or their ability to experience these symptoms. | Animals do not have schizophrenia, as it is a human psychiatric disorder. The context provided discusses various aspects of schizophrenia, such as its symptoms, its impact on patients' lives, and its relation to autism. However, these findings and observations are specific to humans and do not apply to animals. |

| Give me some symptoms of schizophrenia | TF-IDF | Doc2Vec | BERT |
| --- | --- | --- | --- |
| No RAG & No query augmentation | Schizophrenia is a severe mental disorder characterized by a combination of symptoms, including:1. Delusions: Irrational beliefs that are not based on reality.2. Hallucinations: Sensory experiences without external stimuli, such as hearing voices or seeing things.3. Disorganized speech: Incoherent, illogical, or nonsensical speech patterns.4. Disorganized behavior: Unpredictable, bizarre, or inappropriate actions.5. Negative symptoms: Reduced emotional expression, motivation, or initiation of activities.6. Cognitive impairment: Difficulty with attention, memory, and executive functions. These symptoms can vary in intensity and duration, and may require ongoing treatment and support. | | |
| RAG & No query augmentation | Some symptoms of schizophrenia include hallucinations, delusions, disorganized speech and behavior, and negative symptoms such as reduced emotional expression and lack of motivation. The context mentions elevated dopamine receptors in drug-naive schizophrenic patients, suggesting a possible link between dopamine and the development of the disorder. | Some symptoms of schizophrenia include delusions, hallucinations, disorganized speech or behavior, and negative symptoms such as reduced emotional expression or lack of motivation. These symptoms can vary in intensity and may change over time. Diagnosis often involves evaluating the presence and severity of these symptoms, as well as considering the individual's functioning in daily life. | Some symptoms of schizophrenia include:1. Negative symptoms: Flattened affect, poverty of action, and poverty of speech.2. Positive symptoms: Stereotypy, perseveration, and abnormal motor behavior.3. Cognitive impairment and enlarged ventricles, which may reflect abnormal brain structure.4. Associated abnormalities: Perseverative behavior, inappropriate stimulus-elicited behavior, and lack of spontaneous willed behavior. These symptoms are often related to brain areas such as the prefrontal cortex, medial and lateral areas, and the striatum. |
| RAG & query augmentation | Some symptoms of schizophrenia include hallucinations, delusions, disorganized speech and behavior, and negative symptoms such as reduced emotional expression and lack of motivation. Ventricular enlargement, elevated D2 dopamine receptors, and orbital-frontal cortex abnormalities have also been observed in schizophrenic patients. These findings are related to the positive and negative symptoms of the disorder. | Some symptoms of schizophrenia include poverty of speech, monosyllabic answers, and speech that provides little information. Additionally, many schizophrenic patients experience disorganized thinking and planning, as well as hallucinations. The disorder can also lead to difficulties in maintaining employment and an increased risk of suicide attempts. | Some symptoms of schizophrenia include hallucinations, delusions, disorganized speech and behavior, disrupted thought processes, and social withdrawal. These symptoms can vary in intensity and duration, and may be accompanied by changes in mood, energy levels, and overall functioning. It is important to note that these symptoms can be influenced by various factors, such as cognitive processes and environmental factors. |

| How is schizophrenia diagnosed? | TF-IDF | Doc2Vec | BERT |
|---|---|---|---|
| No RAG & No query augmentation | Schizophrenia is diagnosed through a comprehensive evaluation by a mental health professional, considering symptoms, medical history, and the presence of other disorders. The diagnostic criteria from the DSM-5 (Diagnostic and Statistical Manual of Mental Disorders) are typically followed. These include experiencing two or more characteristic symptoms for at least one month, with continuous signs of impaired functioning in areas like work, interpersonal relationships, or self-care. A thorough assessment, including psychiatric interviews, medical tests, and ruling out other causes, is crucial for accurate diagnosis. | | |
| RAG & No query augmentation | Schizophrenia is diagnosed by evaluating the patient's experiences and beliefs, as well as observing symptoms such as hallucinations. The diagnosis requires the patient to report bizarre experiences and beliefs. Although there is a belief that schizophrenia has an organic basis, the diagnosis is not solely dependent on the organic cause. Patients excluded from the diagnosis due to known organic causes may still provide clues about the physiological basis of schizophrenia. The cognitive consequences of brain abnormalities in schizophrenia are often manifested later in life, and understanding the relationship between the mind and the brain is crucial for explaining the symptoms. | Schizophrenia is diagnosed by evaluating a person's signs and symptoms, which can be diverse and complex. Some specific symptoms, like hearing voices discussing them in the third person, are rare but may be encountered in schizophrenic patients. The core symptoms and related disorders can be illustrated using a Venn diagram. Associated abnormalities, such as negative signs, movement disorders, and dementia, are strongly related. Post-mortem evidence shows structural brain changes in schizophrenia. The diagnosis may also involve language analysis and assessing consciousness and information processing. | Schizophrenia is diagnosed through a comprehensive evaluation that includes medical and psychiatric assessments, as well as a thorough examination of the patient's symptoms, history, and functioning. Diagnosis typically involves ruling out other medical conditions and considering various factors such as cognitive impairments, negative and positive symptoms, and the presence of delusions or hallucinations. Psychiatrists and psychologists may use standardized assessments and diagnostic criteria from the Diagnostic and Statistical Manual of Mental Disorders (DSM) or International Classification of Diseases (ICD) to confirm the diagnosis. |
| RAG & query augmentation | Schizophrenia is diagnosed through a comprehensive assessment of symptoms and history, as well as ruling out other possible causes. The diagnostic criteria are usually specified using the Diagnostic and Statistical Manual of Mental Disorders (DSM), such as DSM-III-R. However, different studies may use different diagnostic criteria, and diagnostic criteria alone are not sufficient for interpreting results. Researchers are exploring alternative approaches, such as the Comprehensive Assessment of Symptoms and History (CASH), to better understand and diagnose schizophrenia. | Schizophrenia is diagnosed by evaluating a person's symptoms, medical history, and mental state. The Diagnostic and Statistical Manual of Mental Disorders (DSM-5) and the International Classification of Diseases (ICD-10) provide guidelines for diagnosing schizophrenia. These guidelines include the presence of specific symptoms, such as delusions, hallucinations, disorganized speech, disorganized or catatonic behavior, and negative symptoms. A psychiatrist or psychologist will assess the individual's symptoms, duration, and impact on their daily functioning to make a diagnosis. However, the National Institute of Mental Health has expressed dissatisfaction with a diagnostic approach to research on mental disorders and has proposed alternative research domains, such as cognitive systems and systems for social processes. | Schizophrenia is diagnosed through a combination of evaluating symptoms, medical history, and a thorough psychiatric assessment. Symptoms typically include delusions, hallucinations, disorganized speech and behavior, and negative symptoms such as social withdrawal and lack of motivation. A mental health professional, such as a psychiatrist or psychologist, will assess the individual's symptoms, history, and functioning to determine if the criteria for schizophrenia are met. In some cases, brain imaging or other tests may be conducted to rule out other causes of the symptoms. |

| What role do genetics play in schizophrenia? | TF-IDF | Doc2Vec | BERT |
|---|---|---|---|
| No RAG & No query augmentation | Genetics play a significant role in schizophrenia, as research suggests that about 80% of the risk is heritable. Studies indicate that multiple genes contribute to the development of the disorder, with no single gene responsible for schizophrenia. Environmental factors, such as prenatal exposure to viruses or substance abuse, may also interact with genetic predispositions to increase the risk. However, the complex interplay between genes and environment is not yet fully understood, and more research is needed to uncover the complete picture of schizophrenia's genetic basis. | | |
| RAG & No query augmentation | Genetics play a significant role in schizophrenia, as there is evidence of a strong genetic component in the development of the disorder. Studies have shown that the risk of developing schizophrenia is higher in individuals with a family history of the condition. The dopamine system, which is involved in the control of movement, may also have an indirect role in the control of positive symptoms of schizophrenia, such as cognition and metarepresentational abilities. However, the exact mechanisms and the role of the frontal lobes in schizophrenia are still not fully understood, and further research is needed to uncover the complete picture of the genetic and neurological factors contributing to the disorder. | Genetics play a significant role in schizophrenia, as it is a complex disorder with a strong genetic component. Studies have shown that certain genes are associated with an increased risk of developing schizophrenia. However, the exact role of genetics in schizophrenia is not yet fully understood, as it likely involves a combination of genetic and environmental factors. The context provided discusses the importance of emotional expression and the ability to understand and respond to others' mental states, which is impaired in autistic and schizophrenic individuals. The dopamine system, which is involved in movement control, may also play a role in the positive symptoms of schizophrenia. Structural brain changes have been observed in individuals with schizophrenia, and these changes may contribute to the development of the disorder. Leslie's work on mentalizing and metarepresentational abilities has provided insights into the cognitive mechanisms involved in schizophrenia, particularly in relation to pretend play and understanding the mental states of others. In summary, genetics, dopamine system, and brain structure are all factors that play a role in schizophrenia, and further research is needed to fully understand their contributions to the development and progression of the disorder. | Genetics play a significant role in schizophrenia, as research suggests that genetic factors contribute to the development of the disorder. Studies have identified multiple genes associated with schizophrenia, although the exact combination of genetic variations and environmental factors that lead to the condition remains complex and not fully understood. |
| RAG & query augmentation | Genetics play a significant role in schizophrenia, as there are numerous 'risk' genes associated with the disorder. However, no specific or reliable markers have been found yet. Studies of the brain in schizophrenia have shown increased dopamine-receptor sensitivity, which may contribute to the development of symptoms. The cognitive approach, which focuses on specific symptoms rather than the diagnostic category, can help better understand the role of genetics in schizophrenia. | Genetics play a significant role in schizophrenia, as they can provide clues to the identity of the final common pathway in the brain that functions abnormally. While the exact cause of schizophrenia is not fully understood, it is believed that a combination of genetic, environmental, and cognitive factors contributes to the development of the disorder. Studies have shown that schizophrenic patients can differ widely in their symptoms, making it challenging to draw conclusions from early studies based solely on a hospital diagnosis. Recent research has focused on specifying diagnostic criteria and examining the cognitive aspects of schizophrenia, such as reasoning, perseveration, distractibility, and incoherent behavior. | Genetics play a significant role in schizophrenia, as research suggests that both genetic and environmental factors contribute to the development of the disorder. Studies have identified multiple genes associated with schizophrenia, and family studies have shown a higher risk of developing the condition among individuals with affected relatives. However, the exact combination of genetic and environmental factors that lead to schizophrenia is still not fully understood. In the context of amphetamine's effects on behavior, it is evident that genetics may influence an individual's susceptibility to experiencing schizophrenic symptoms, such as delusions and auditory hallucinations, when exposed to the drug. |

| How does schizophrenia impact a person's daily life? | TF-IDF | Doc2Vec | BERT |
|---|---|---|---|
| No RAG & No query augmentation | Schizophrenia affects a person's daily life in various ways. It can cause hallucinations, delusions, and disorganized thinking, leading to difficulty in social interactions, work or school performance, and self-care. People with schizophrenia may experience emotional instability, low motivation, and reduced ability to cope with stress. Treatment, including medication and therapy, can help manage symptoms and improve daily functioning. | | |
| RAG & No query augmentation | Schizophrenia impacts a person's daily life by affecting their cognitive abilities, emotions, and social interactions. Symptoms include difficulty recognizing familiar people, reduced emotional responses, and confusion between reality and delusions. The exact cause of these symptoms is not yet fully understood, and researchers are exploring various approaches, such as computational neuropsychiatry, to better understand the disorder. Explaining schizophrenia involves identifying its causes, which can help develop more effective treatments and improve the quality of life for those affected. | Schizophrenia impacts a person's daily life by affecting their thoughts, emotions, and behaviors. It can lead to difficulties in understanding mental states, agency, and social interactions. The cognitive approach to understanding the symptoms of schizophrenia has been influenced by various developments, such as computational neuropsychiatry. This field aims to relate cognitive processes to discrete brain systems, which could help explain the disorder's manifestation in the brain. However, the impact of the cognitive approach has been reduced due to the availability of structural and functional brain imaging, which allows for direct observation of brain-behavior relationships and diagnosis. This shift in focus has dominated the study of schizophrenia, particularly in the context of genetics, drug treatment, and brain imaging. | Schizophrenia impacts a person's daily life by affecting their ability to understand and communicate with others. This is due to the failure of meta representation, which is the ability to infer the mental states of others. Adult schizophrenic patients, unlike autistic children, continue to make inferences about others' mental states but often get them wrong. This can be observed in their speech and non-verbal communication. Schizophrenic patients may experience reduced action, poverty of speech, and incoherent discourse. At a lower level of severity, they may have inner speech and thought that are experienced as verbal hallucinations. At a higher level of severity, these thoughts are spoken and mixed up with vocal speech. This can lead to confusion for both the patient and the listener. In addition, schizophrenic patients may exhibit stereotyped and perseverative behavior in their speech and other acts. Their discourse is often not structured in a way that allows the listener to understand how the components link together or the purpose of the communication. This can make daily interactions and communication challenging for individuals with schizophrenia. |
| RAG & query augmentation | Schizophrenia impacts a person's daily life by causing positive and negative symptoms, affecting cognitive processing and social functioning. Positive symptoms include hallucinations and delusions, while negative symptoms involve reduced emotional expression and motivation. These symptoms can lead to difficulties in communication, relationships, and self-care. Additionally, abnormalities in the experience of action and faulty awareness of others' intentions contribute to the challenges faced by individuals with schizophrenia. Understanding these symptoms and their underlying causes can help develop better treatments and support systems for those affected. | Schizophrenia impacts a person's daily life by affecting their cognitive abilities, such as language, memory, and decision-making. It can lead to difficulties in understanding irony, processing context, and choosing appropriate words. Additionally, schizophrenia may cause hallucinations and delusions, which can interfere with a person's ability to function normally in their daily environment. Researchers are working to understand the link between psychological abnormalities and brain dysfunction in schizophrenic patients, in hopes of finding better treatments and improving their quality of life. | Schizophrenia impacts a person's daily life by affecting their behavior, cognition, and emotions. Patients with frontal lobe lesions, which are often associated with schizophrenia, exhibit perseverative behavior, inappropriate stimulus-elicited behavior, and lack of spontaneous willed behavior. They also struggle with long-term planning, making it difficult to manage daily tasks. Positive symptoms, such as hallucinations and delusions, are common in schizophrenia. These symptoms can be challenging to classify, as they involve false perceptions and false beliefs. The reduction in fluency in schizophrenic patients is not due to a limited vocabulary, but rather their difficulty in spontaneously producing words in a given category. Additionally, schizophrenic patients often have reduced syntactic complexity in their speech, with fewer relative clauses, less clausal embedding, and shorter utterances. The experiences of hallucinations in schizophrenia are often perceived as "real" perceptions, and there is a belief that a form of self-monitoring plays a vital role in modifying our perception of the world. Defects in self-monitoring processes might lead to the experience of hallucinations, as suggested by Ralph Hoffman (1986). Incoherent speech may also be linked to hallucinations, as proposed by Ralf Hoffman (1986). |

| Is there a link between substance abuse and schizophrenia? | TF-IDF | Doc2Vec | BERT |
|---|---|---|---|
| No RAG & No query augmentation | Yes, there is a link between substance abuse and schizophrenia. Substance abuse can trigger or exacerbate symptoms in individuals with schizophrenia, while schizophrenia can increase the risk of substance abuse. Both factors can contribute to a cycle of negative outcomes, such as cognitive impairment, social problems, and reduced life expectancy. | | |
| RAG & No query augmentation | There is a potential link between substance abuse and schizophrenia, as some authors have suggested that defects in self-monitoring processes might lead to hallucinations. Ralf Hoffman (1986) proposed that incoherent speech in schizophrenic patients might be due to unrelated words and phrases being inserted randomly into their speech. Additionally, Parkinson's disease, which can be treated with drugs that stimulate the dopamine system, has been linked to schizophrenia. However, the cognitive basis of these formulations is vague, and further research is needed to establish a clear connection between substance abuse and schizophrenia. | There is a potential link between substance abuse and schizophrenia, as suggested by the context. Some authors propose that defects in self-monitoring processes might lead to hallucinations. Ralf Hoffman (1986) also suggested a link between incoherent speech and hallucinations, stating that schizophrenic speech appears incoherent due to unrelated words and phrases being inserted randomly. Post-mortem evidence of structural brain changes in schizophrenia further supports the connection between the brain and the disorder. However, there is still a need for more research and advancements in cognitive science and neurophysiology to fully understand the relationship between substance abuse, schizophrenia, and the mind-brain distinction. | There is a link between substance abuse and schizophrenia, as substance abuse can trigger or exacerbate psychotic symptoms in individuals with schizophrenia. The context provided discusses antipsychotic drugs and their effectiveness in managing schizophrenia by blocking dopamine receptors. It also mentions studies on inner experiences and contention scheduling in the brain, which can be affected by substances like amphetamine. These findings suggest a connection between substance abuse and the worsening or onset of schizophrenia symptoms. |
| RAG & query augmentation | There is a potential link between substance abuse and schizophrenia, as substance abuse can have an organic basis, which may contribute to the development of schizophrenia. However, the relationship is complex and not fully understood. Some studies suggest that increased dopamine-receptor sensitivity in schizophrenia may be related to the effects of certain drugs, such as amphetamine. Additionally, the use of certain medications for schizophrenia can have parkinsonian side-effects, which may be similar to the effects of some drugs. It is important to note that the link between substance abuse and schizophrenia is not definitive, and further research is needed to establish a clear connection. | There is a potential link between substance abuse and schizophrenia, as evidenced by the effects of amphetamine on animals, which can produce behavioral abnormalities similar to those in schizophrenia. However, the cognitive basis of these formulations is too vague to be useful, and the connection is mainly established through task performance rather than underlying cognitive processes. This approach is based on behaviorism and is essentially an argument by analogy. Further research is needed to better understand the relationship between substance abuse and schizophrenia. | There is a potential link between substance abuse and schizophrenia, as receptors have been found to be increased in drug-naïve schizophrenic patients in the USA. However, in Sweden, receptor sensitivity has been found to be normal. The relationship between substance abuse and schizophrenia is complex and not fully understood, as there is also indirect evidence against the simple dopamine hypothesis. Behavioral abnormalities in schizophrenia, such as incoherence, incongruity, and stereotyped behavior, may be related to the brain's contention scheduling, which is instantiated in the basal ganglia. Amphetamine's effects on contention scheduling could potentially explain some positive behavioral abnormalities in schizophrenia, but there is a major problem in linking animal studies to human behavior due to differences in the effects of amphetamine on animals and humans. |

| Is there a link between substance abuse and schizophrenia? (with extra article) | TF-IDF | Doc2Vec | BERT |
|---|---|---|---|
| No RAG & No query augmentation | Yes, there is a link between substance abuse and schizophrenia. Substance abuse can trigger or exacerbate symptoms in individuals with schizophrenia, while schizophrenia can increase the risk of substance abuse. Both factors can contribute to a cycle of negative outcomes, such as cognitive impairment, social problems, and reduced life expectancy. | | |
| RAG & No query augmentation | There is a potential link between substance abuse and schizophrenia, as comorbid substance abuse has been found in recent-onset schizophrenia. Parkinson's disease, which can be treated with drugs that stimulate the dopamine system, has also been linked to schizophrenia. However, the cognitive basis of these connections is vague and further research is needed to establish a clear relationship between substance abuse and schizophrenia. | There is a potential link between substance abuse and schizophrenia, as suggested by various studies and authors. Defects in self-monitoring processes might lead to hallucinations, and incoherent speech has been proposed as a possible indicator of schizophrenia. Studies have also shown that substance use and abuse are prevalent in first-episode psychosis and can influence the onset and course of schizophrenia. Additionally, there is evidence of lower negative symptom scores among cannabis-dependent patients with schizophrenia-spectrum disorders. However, more research is needed to establish a definitive connection between substance abuse and schizophrenia. | There is a potential link between substance abuse and schizophrenia, as evidenced by the effects of amphetamine on animals. Amphetamine disrupts contention scheduling in the basal ganglia, leading to behavioral abnormalities such as incoherence, incongruity, and stereotyped behavior. However, more systematic studies are needed to examine these hypotheses, as current research often relies on comparing "schizophrenics" with haphazardly defined control groups. |
| RAG & query augmentation | Yes, there is a link between substance abuse and schizophrenia. Multiple studies have shown that individuals with schizophrenia are more likely to have a history of substance abuse, and substance abuse can exacerbate symptoms and impact outcomes. Research has also indicated that comorbidity of substance abuse and severe mental illness is common, with prevalence and consequences varying depending on the type of substance and other factors. | There is a link between substance abuse and schizophrenia, as evidenced by several studies. These studies show that individuals with severe mental illness, such as schizophrenia, have a higher prevalence of substance abuse. Additionally, substance abuse can precipitate and influence the onset and course of schizophrenia. Motivation to quit using substances among individuals with schizophrenia is also an important factor to consider in treatment models. | There is a potential link between substance abuse and schizophrenia, as substance abuse can increase the risk of developing schizophrenia or exacerbate its symptoms. However, the exact relationship between the two is not fully understood. Some studies suggest that substance abuse may contribute to the development of schizophrenia by altering brain chemistry or triggering a psychotic episode in individuals who are genetically predisposed. On the other hand, substance abuse can also be a symptom or consequence of untreated or poorly managed schizophrenia, as individuals may turn to substances as a coping mechanism or to self-medicate for negative symptoms. It is important to address both substance abuse and schizophrenia through appropriate treatment and support to improve overall mental health and well-being. |

| Is bipolarity similar to schizophrenia | TF-IDF | Doc2Vec | BERT |
|---|---|---|---|
| No RAG & No query augmentation | Bipolarity, also known as bipolar disorder, is a mental health condition characterized by extreme mood swings between mania and depression. Schizophrenia, on the other hand, is a severe and chronic mental disorder involving distorted thoughts, perceptions, and emotions. While both conditions involve mental health challenges, they are distinct and not similar in terms of symptoms, causes, and treatments. | | |
| RAG & No query augmentation | Bipolarity and schizophrenia are different mental health conditions, although they share some similarities. Bipolar disorder is characterized by extreme mood swings between mania and depression, while schizophrenia involves symptoms such as hallucinations, delusions, and disorganized thinking. Both conditions can involve difficulties in self-monitoring and agency, but they have distinct causes and treatments. Bipolarity is not similar to schizophrenia but understanding the similarities and differences between them can help in developing better treatments and support for those affected by these conditions. | Bipolarity and schizophrenia are different mental health conditions, although they share some similarities. Bipolar disorder is characterized by extreme mood swings between mania and depression, while schizophrenia involves symptoms such as hallucinations, delusions, and disorganized thinking. Although both conditions can involve problems in language and pragmatics, they are not the same. Schizophrenia is a more complex and chronic disorder, and its diagnosis is based on a combination of symptoms, not just language issues. | Bipolarity and schizophrenia are different mental health conditions, although they share some similarities. Bipolar disorder is characterized by extreme mood swings between mania and depression, while schizophrenia involves symptoms such as hallucinations, delusions, and disorganized thinking. Although both conditions can involve cognitive impairments, they have distinct diagnostic criteria and treatment approaches. The context provided discusses various studies and research on brain damage, cognitive impairments, and symptoms in schizophrenia, but does not directly compare bipolarity and schizophrenia. |
| RAG & query augmentation | Bipolarity and schizophrenia are different mental health conditions, although they share some similarities. Bipolar disorder is characterized by alternating periods of mania and depression, while schizophrenia involves symptoms such as delusions, hallucinations, and disorganized thinking. Although depression can be a risk factor for both conditions, they have distinct underlying causes and treatments. Research on animals and the study of different types of psychoses have helped distinguish between these two disorders. | Bipolarity and schizophrenia are different mental health conditions, although they share some similarities. Bipolar disorder is characterized by extreme mood swings between mania and depression, while schizophrenia involves disruptions in thought processes, perceptions, and emotional responsiveness. While both conditions can involve delusions and hallucinations, bipolar disorder is primarily a mood disorder, while schizophrenia is considered a psychotic disorder. It is essential to consult a mental health professional for accurate diagnosis and treatment. | Bipolarity and schizophrenia are different mental health conditions, although they share some overlapping symptoms. Schizophrenia is characterized by hallucinations, delusions, and negative symptoms, such as social withdrawal and flattened emotions. Bipolar disorder, on the other hand, involves episodes of mania and depression, with mood swings being the primary concern. While both conditions can include hallucinations and delusions, bipolarity does not typically involve the same level of cognitive impairment or negative symptoms as schizophrenia. It is essential to consult a mental health professional for accurate diagnosis and treatment. |

| What is the best medication | TF-IDF | Doc2Vec | BERT |
|---|---|---|---|
| No RAG & No query augmentation | The best medication depends on the specific condition, individual needs, and potential side effects. It is essential to consult a healthcare professional for proper diagnosis and recommendations. They can guide you towards the most suitable medication based on your medical history and current health status. | | |
| RAG & No query augmentation | The best medication for schizophrenia and psychoses of uncertain etiology depends on the individual's specific symptoms and needs. Neuroleptic and anticholinergic medications can help manage positive symptoms, such as hallucinations and delusions, by reducing the activity of dopamine in the brain. However, it is essential to consider the patient's cognitive function and any neuroendocrine changes that may occur due to the medication. In some cases, a combination of medication and physical treatment may be more effective. It is crucial to consult with a healthcare professional to determine the most suitable treatment plan for each patient. | The best medication for schizophrenia depends on the individual's symptoms and needs. Antipsychotic medications, such as olanzapine, risperidone, and clozapine, are commonly used to treat positive symptoms like delusions and hallucinations. However, these medications can also affect cognitive function, making it more difficult for patients to think and act spontaneously. It is essential to consider the trade-offs between symptom control and cognitive function when choosing the most appropriate medication for each patient. | The best medication for schizophrenia depends on the individual's symptoms and response to treatment. Antipsychotic drugs, such as olanzapine, risperidone, and clozapine, are commonly prescribed to manage symptoms like hallucinations, delusions, and disorganized thinking. Other medications, like mood stabilizers or antidepressants, may also be used in combination with antipsychotics. It is essential to work closely with a healthcare professional to find the most suitable medication and dosage for each patient. |
| RAG & query augmentation | The best medication for schizophrenia depends on the individual's symptoms and the severity of their condition. In the context provided, alpha-flupenthixol showed a significantly greater reduction in symptom severity compared to beta-flupenthixol and placebo after two weeks of treatment. This confirms that dopamine blockade is a necessary condition for reducing symptom severity. It is important to note that medication and treatment options may vary, and it is crucial to consult with a healthcare professional for personalized recommendations. | The best medication for bipolar disorder depends on the individual's specific symptoms and needs. However, some commonly used medications include mood stabilizers like lithium, valproate, and carbamazepine, as well as antipsychotics for psychotic episodes. Antidepressants may also be prescribed, but they should be used with caution as they can trigger manic episodes in some patients. It is essential to consult with a healthcare professional to determine the most appropriate medication and dosage for each person's unique situation. | The best medication for schizophrenia depends on the individual's symptoms and medical history. Antipsychotic drugs, such as olanzapine, risperidone, and clozapine, are commonly prescribed to manage symptoms like delusions, hallucinations, and disorganized thinking. It is essential to work closely with a healthcare professional to find the most suitable medication and dosage for each patient. |